\documentclass[
reprint,
 superscriptaddress,
 amsmath,amssymb,
 aps,
 cleveref,
]{revtex4-1}
\usepackage{fullpage}
\usepackage{setspace}
\usepackage{amssymb}
\usepackage{hyperref}
\usepackage{mathtools}

\usepackage{algorithm}
\usepackage{algpseudocode}

\usepackage{algcompatible}

\DeclarePairedDelimiter\abs{\lvert}{\rvert}%
\DeclarePairedDelimiter\norm{\lVert}{\rVert}%

\makeatletter
\let\oldabs\abs
\def\abs{\@ifstar{\oldabs}{\oldabs*}}
\let\oldnorm\norm
\def\norm{\@ifstar{\oldnorm}{\oldnorm*}}
\makeatother
\def\BState{\State\hskip-\ALG@thistlm}

\newcommand{\defeq}{\vcentcolon=}

\DeclareMathOperator*{\argmax}{arg\,max}

\begin{document}

\title{Efficient Phase Diagram Sampling by Active Learning}
\author{Chengyu Dai}%
\affiliation{%
Department of Physics, University of Michigan, Ann Arbor, Michigan 48109, USA \\
}%
\affiliation{%
Biointerfaces Institute, University of Michigan, Ann Arbor, Michigan 48109, USA\\
}%
\author{Isaac R. Bruss}%
\affiliation{%
Biointerfaces Institute, University of Michigan, Ann Arbor, Michigan 48109, USA\\
}%
\affiliation{%
Department of Chemical Engineering, University of Michigan, Ann Arbor, Michigan 48109, USA\\
}%
\author{Sharon C. Glotzer}%
 \email{sglotzer@umich.edu}
\affiliation{%
Department of Physics, University of Michigan, Ann Arbor, Michigan 48109, USA \\
}%
\affiliation{%
Biointerfaces Institute, University of Michigan, Ann Arbor, Michigan 48109, USA\\
}%
\affiliation{%
Department of Chemical Engineering, University of Michigan, Ann Arbor, Michigan 48109, USA\\
}%
\affiliation{%
Department of Materials Science and Engineering, University of Michigan, Ann Arbor, Michigan 48109, USA\\
}%

\date{\today}% It is always \today, today,

\begin{abstract}

We address the problem of efficient phase diagram sampling by adopting active learning techniques from machine learning, and achieve an 80\% reduction in the sample size (number of sampled statepoints) needed to establish the phase boundary up to a given precision in example application. Traditionally, data is collected on a uniform grid of predetermined statepoints. This approach, also known as grid search in the machine learning community, suffers from low efficiency by sampling statepoints that provide no information about the phase boundaries. We propose an active learning approach to overcome this deficiency by adaptively choosing the next most informative statepoint(s) every round. This is done by interpolating the sampled statepoints' phases by Gaussian Process regression. An acquisition function quantifies the informativeness of possible next statepoints, maximizing the information content in each subsequently sampled statepoint. We also generalize our approach with state-of-the-art batch sampling techniques to better utilize parallel computing resources. We demonstrate the usefulness of our approach in a few example simulations relevant to soft matter physics, although our algorithms are general. Our active learning enhanced phase diagram sampling method greatly accelerates research and opens up opportunities for extra-large scale exploration of a wide range of phase diagrams by simulations or experiments.

\end{abstract}

\maketitle

% \pagebreak

%--INTRODUCTION--%
\section{Introduction}

Mapping phase diagrams is a central task in many branches of both physical and chemical sciences.  Phase diagrams map macroscopic, thermodynamic phases onto statepoint variables such as pressure, density, temperature, and concentration. Alchemical parameters can be varied to produce ``phase" diagrams mapped onto stoichiometry, interaction range, patchiness, and even particle shape. The phase boundaries separating macroscopic phases are of particular importance to studies of phase transitions and phase coexistence. Despite its ubiquitous utility, constructing phase diagrams - whether by experiments or simulations -- remains a laborious and costly task, even for simple systems with only a few thermodynamic phases. Traditionally, researchers conduct experiments or simulations at statepoints uniformly laid on a grid \textit{a priori}, as the one shown in Fig. \ref{example}(a). While this grid search method is simple to understand and implement, it is inefficient because the scheduling of later experiments/simulations ignores the valuable feedback from previous experiments/simulations. In realistic scenarios relevant to the chemical and physical sciences, the overhead to implement and run a more advanced sampling algorithm is usually negligible compared to sampling an extra statepoint. Thus, an advanced sampling strategy is highly desirable.

With recent developments in machine learning techniques \cite{kapoor2010gaussian, gonzalez2016batch, azimi2010batch, wu2016parallel, wang2016parallel, kathuria2016batched}, we now have a powerful arsenal to develop an advanced sampling strategy.  Problems similar to the construction of phase diagrams arise naturally in the field of active learning \cite{dasgupta2005analysis, castro2008active, dasgupta2008general, kapoor2010gaussian}, where an algorithm actively queries a particular subset of labeled data rather than passively using a predetermined training dataset. Related techniques were also developed for Bayesian optimization in operations research \cite{rasmussen2006gaussian,gonzalez2016batch,wu2016parallel, wang2016parallel}. Combining some of the latest advancements in both fields, we present an active learning framework for efficient phase boundary sampling. By sequentially sampling only the most informative statepoints, our approach to finding a phase boundary reduces the number of statepoints needing to be sampled to only 20\% of the number of statepoints needed in the conventional grid search method, while maintaining roughly the same precision on the location of the phase boundaries. The high efficiency thus opens up opportunities for extra-large scale phase diagram explorations that would otherwise be prohibitive. Moreover, our approach can take advantage of modern parallel experiment or simulation resources by supporting batch sampling, where the algorithm returns multiple statepoints for phase evaluation at each round.

\section{Active Learning Framework}

\begin{figure}
\centering
\includegraphics[width=0.45\textwidth]{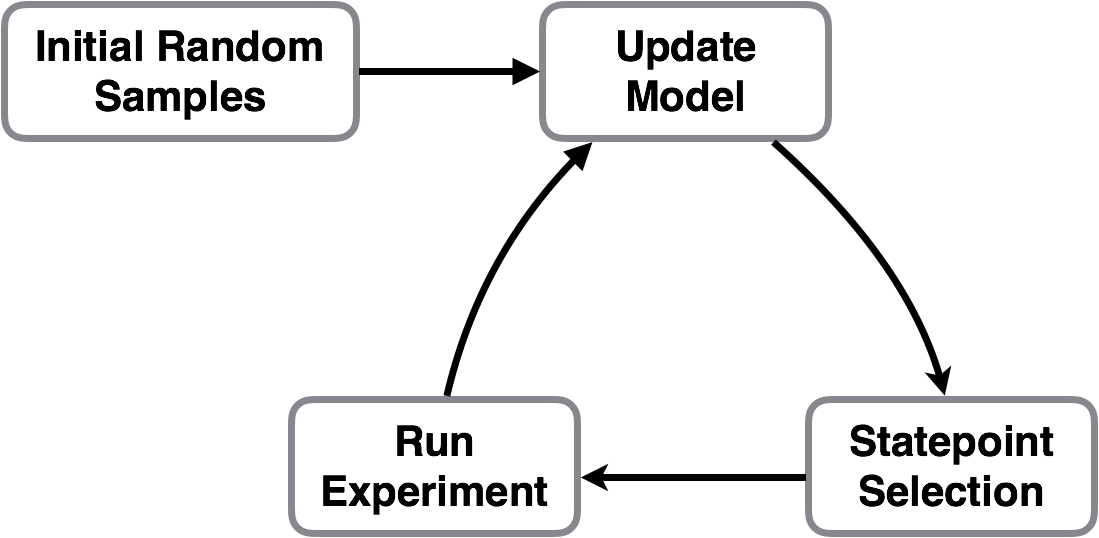}
\caption{A schematic that summarizes the active learning framework.}
\label{schematics}
\end{figure}

We begin by posing our problem in an active learning setting by connecting related concepts in chemical physics and machine learning, see Table \ref{terminology}. Without loss of generality, in this paper we limit our discussion to the case where only two distinct phases are present in the phase diagram, and denote the phase label to be $y$. The phases might be, for example, liquid and gas, polymer-rich and polymer-poor, face-centered cubic and body-centered cubic, etc.  We consider statepoints with $d$ parameters as a vector $\mathbf{x}$ in $d$-dimensional vector space $\mathbb{R}^d$. Our goal is to perform the phase evaluation (i.e. run costly experiments or simulations) only at the most informative trial statepoints to improve sampling efficiency. First, we use an \textbf{interpolation method} that can be fitted with the phases from the observed statepoints to the entire parameter space of interest. Next, we define an \textbf{acquisition function} $\alpha(\mathbf{x})$ to quantify the ``informativeness" of a trial point based on the uncertainty in the outcome of its phase evaluation and whether it promises to add more precision to the evaluations of the phase boundary. We perform phase evaluations at the trial statepoint with the current highest acquisition function value. Finally, we repeat the above operations, each round interpolating again with the additional data, until reaching the target number of iterations or the desired precision. The simple schematic in Fig. \ref{schematics} summarizes the active learning framework applied to phase diagram evaluation, and a formal sequential algorithm is given in Algorithm \ref{algoseq}, with the detailed derivation of each step explained in the following sections.

\bigskip
\begin{figure*}
\begin{minipage}{\linewidth}
\begin{algorithm}[H]
  \caption{Active Learning Based Phase Diagram Sampling - Sequential}
  \label{algoseq}
  
\begin{flushleft}
   \textbf{Input:} {Randomly sampled initial dataset $\mathcal{D}_1 \defeq \left\{ \mathbf{x}_i, y_i \right\} ^n_{i=1}$ that contains statepoints in both phases, maximum iteration $m$ } \end{flushleft}
  \begin{algorithmic}
	\FOR{$t=1$ to $m$}
      \STATE {Fit a Gaussian Process to $\mathcal{D}_t$ and compute the acquisition function $\alpha(\mathbf{x})$ \;}
      \STATE {Compute $\mathbf{x}^\star \leftarrow \argmax (\alpha(\mathbf{x})$ by gradient ascent method \;}
      \COMMENT {Select trial point by maximizing acquisition function}
      \STATE {Run experiment or simulation to evaluate the phase $y^\star$ at trial point $\mathbf{x^\star}$\;}
      \STATE {$\mathcal{D}_{t+1} \leftarrow \mathcal{D}_{t} \cup \left\{ (\mathbf{x}^\star, y^\star)\right\}$ \;}
      \COMMENT {Add observed statepoint in this round into the observed set}
    \ENDFOR
  \end{algorithmic}
\end{algorithm}
\end{minipage}
\end{figure*}

\section{Model for Interpolation}

At each iteration in the active learning process, we need to update a model for interpolating the phase function from the observed statepoints to the entire space of interest. The model is an essential part of the active learning framework, because to quantify the information gain in choosing an as-yet-unsampled trial statepoint, we must first estimate the possible outcomes of the experiment. Our assumption about the phase diagram is that the phase of a statepoint is highly correlated with the phase of its neighboring statepoints, because experiments with similar parameters are likely to find similar results. Such correlation is exactly why active learning performs better than a grid search, because active learning uses an interpolation model to utilize this information and chooses fewer redundant statepoints to sample.

A natural model for pure interpolation should be able to capture the spatial correlation of the phase function and provide estimates of not-sampled statepoints' phases. According to these criteria, we choose Gaussian Process (GP) regression \cite{rasmussen2006gaussian}, a well studied and widely adopted method in research and industry. In our setting where only two phases are present, a regression method is equivalent to solving a classification problem relaxed to real values. GP regression has been widely used to model data with spatial correlation due to a unique advantage of providing probabilistic interpolation estimates where the uncertainty of output estimates can also be quantified \cite{murphy2012machine}. The uncertainty for interpolation also provides important information for the active learning framework to choose the next trial point.

To formalize the GP interpolation, we are given a set of statepoints $\mathbf{X} = \{\mathbf{x}_1, ..., \mathbf{x}_n\}$ from past trials whose phases have already been evaluated as $\mathbf{y}=\{y_1, ..., y_n\}$. Without loss of generality, the phase labels can be denoted as $y_i \in \{-1,1\}$. Following the derivation developed in \cite{rasmussen2006gaussian}, GP interpolation assumes that the joint distribution of sampled labels $\mathbf{y}$ and prediction labels $\mathbf{y^\star}$ follows the Gaussian prior distribution,

\begin{equation}
  \left [ \begin{array}{lcr}
  \mathbf{y} \\ \mathbf{y^\star}
  \end{array}  \right ] 
  \sim \mathcal{N} \left ( \mathbf{0}, 
  \left [ \begin{array}{lcr}
  K(X,X) & K(X,X^\star) \\ K(X^\star,X) & K(X^\star,X^\star)
  \end{array}  \right ] 
  \right ).
  \label{prior}
\end{equation}
The kernel matrix $K(X,X^\star)$ represents the covariances between all pairs of sampled and unknown statepoint labels. We define $K(X,X), K(X^\star,X)$, and $K(X^\star,X^\star)$ similarly. The kernel matrices are constructed such that the covariances decrease as the distance between statepoints grows. Representing the nonnegative pairwise covariance, the kernel matrix elements are assumed to be given by a positive-definite kernel function $K_{ij} = k(\mathbf{x_i},\mathbf{x_j})$. We adopt the most common choice of kernel function that satisfies all above requirements, the squared exponential (SE) function \cite{rasmussen2006gaussian, wilson2013gaussian}, also known as a Gaussian kernel, 
\begin{equation*}
	\mathrm{cov}(y(\mathbf{x_i}),y(\mathbf{x_j})) = k(\mathbf{x_i}, \mathbf{x_j}) = \exp{(-\frac{1}{2} \norm{\frac{\mathbf{x_i} - \mathbf{x_j}}{2l^2}}^2)}.
\end{equation*}
where $l$ represents the characteristic length scale. $l$ can be learned from the data by automatic relevance determination (ARD) according to the standard practice of Bayesian Optimization research \cite{snoek2012practical}.

With these assumptions, the interpolated (predicted) phase function for the unknown statepoints follows the conditional distribution, calculated by updating the prior probability distribution (Eq. \ref{prior}) after observing sampled statepoints' phases. Thus $\mathbf{y^\star} \sim \mathcal{N}(\mathbf{\mu}_{\mathbf{y}}^\star, \mathrm{var}(\mathbf{y^\star}))$, i.e. the unknown statepoints' phase function follows a Gaussian distribution, where the mean ($\mu_{\mathbf{y^\star}}$) and variance ($\mathrm{var}({\mathbf{y^\star}})$) is given by 
\begin{widetext}
\begin{equation}
	\mathbf{\mu}_{\mathbf{y^\star}} \defeq \mathbb{E}\left[ \mathbf{y^\star} | \mathbf{y},X,X^\star \right] = K(X^\star, X)\left[ K(X,X)\right]^{-1}\mathbf{y},
\end{equation}
\begin{equation}
	 \mathrm{var}(\mathbf{y}^\star) = K(X^\star, X^\star) - K(X^\star, X)\left[ K(X,X)\right]^{-1}K(X,X^\star).
\end{equation}
\end{widetext}
First, the interpolation estimate is probabilistic rather than deterministic because what we calculate is the distribution, fully characterized by the mean ($\mu_{\mathbf{y^\star}}$) and variance ($\mathrm{var}({\mathbf{y^\star}})$) of the distribution, both of which can be analytically expressed. Although the actual phase evaluation returns only discrete phase labels $y_i \in \{ -1,1 \}$, the interpolation predictions are relaxed to real values that may occasionally be outside of the range $[-1,1]$. In particular, a $0$ value of predictive mean $\bar{\mathbf{y}}^\star$ shows that the GP interpolation algorithm considers this statepoint as lying exactly on the estimated phase boundary based on all observed information. Meanwhile, the variance of the prediction $\mathrm{var}(\mathbf{y}^\star)$ quantifies the uncertainty of the interpolation estimate. While it is possible to use more complicated interpolation methods, we choose GP in favor of its interpretability and simplicity.

\begin{figure*}
\centering
\includegraphics[width=1\textwidth]{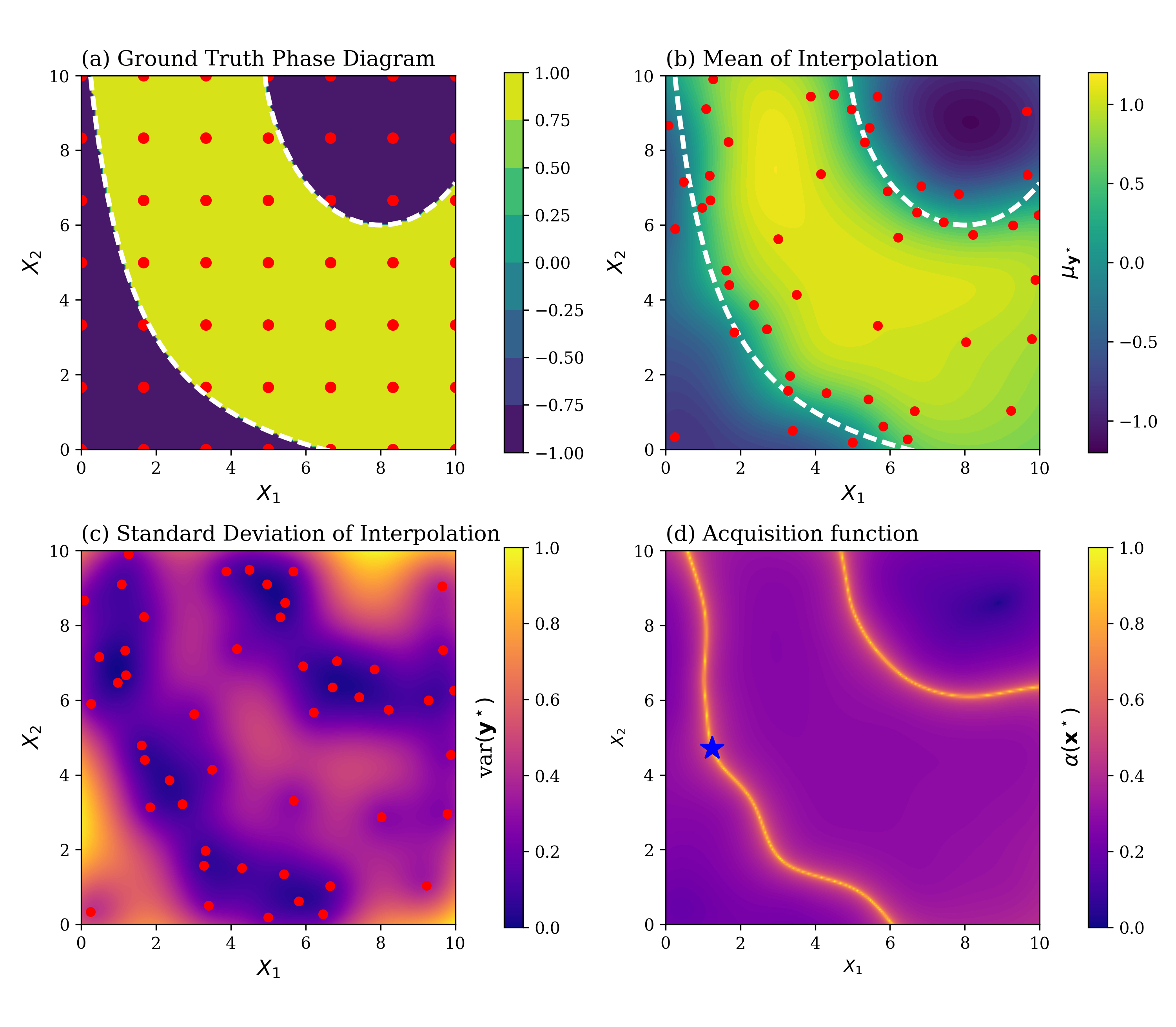}
\caption{An example of the Gaussian Process (GP) interpolation model and acquisition function at a given iteration for a synthetic phase diagram in two dimensions. (a) The ground truth (i.e. actual) phase boundary, with a 7-by-7 grid plotted on top to illustrate the grid search method. (b) The expected phase value calculated by the GP interpolation. Red dots represent observed statepoints. (c) The standard deviation (uncertainty) of the GP interpolation estimates to the phase function. (d) The acquisition function value calculated according to Eq. \ref{t_stat}. The value is rescaled to be in the range between $0$ and $1$. The star represents the statepoint with the maximum acquisition function value, i.e. the selected trial statepoint for the next experiment.}
\label{example}
\end{figure*}

To demonstrate the interpolation model, a synthetic example is used. Note that our algorithm does not depend on the assumption that each phase region is connected, as shown in our example ground truth (that is, actual) phase diagram in Fig. \ref{example}(a). Fig. \ref{example}(b) shows the mean value of the GP interpolation is very close to the ground truth even with only 50 sampled statepoints. Moreover, the uncertainty of the interpolation is quantified in the model by the standard deviation, shown in Fig. \ref{example}(c). Since GP utilizes the spatial correlation between statepoints, the regions with fewer sampled statepoints would see larger uncertainty in the estimate of their proximity.

\begin{table}[]
\centering
\caption{Terminology matching between phase boundary mapping and classifier training}
\medskip
\label{terminology}
\begin{tabular}{|l|l|}
\hline
interpolating phase diagram     & training classifier \\ \hline
statepoint             			& feature vector      \\ \hline
phase                  			& class               \\ \hline
phase boundary         			& decision boundary   \\ \hline
distance to phase boundary	    & margin			  \\ \hline
\end{tabular}
\end{table}

\section{Acquisition Function}
To quantify the reward when the next sample to be evaluated is placed at $\mathbf{x}$, we propose the acquisition function $\alpha(\mathbf{x})$ as a function of a statepoint's interpolated phase estimates $\bar{\mathbf{y}}^\star$ and $\mathrm{var}(\mathbf{y}^\star)$ (so also a function of $\mathbf{x}$ indirectly). The terminology ``acquisition function'' is a standard term quantifying the ``informativeness'' of a potential trial statepoint borrowed from the related field of Bayesian Optimization \cite{snoek2012practical}. Thus, given the functional form of $\alpha(\mathbf{x})$, the algorithm for identifying the next state point is simply choosing the statepoint that yields the largest acquisition function value, $\tilde{\mathbf{x}} = \argmax{(\alpha(\mathbf{x}))}$.

The design of the acquisition function needs to balance the trade-off between minimizing uncertainty by gathering diverse samples (known as ``exploration'') and increasing precision on phase boundary regions (known as ``exploitation''). With GP as our interpolation method, one candidate acquisition function based on exploration is to choose $\sqrt{\mathrm{var}(y^\star) + \sigma^2}$, where $\sigma$ denotes the magnitude of the error in the phase evaluation from experiment or simulation. This functional form intuitively states that the statepoint with the largest value of estimated variance should be sampled first. Another candidate functional form based on exploitation is $-\left| \bar{\mathbf{y}}^\star \right|$. The largest possible value of this candidate function is $0$, so this choice will guide the algorithm to sample statepoints at the estimated phase boundary, ``exploiting'' the current knowledge of the phase diagram and refining the phase boundary regions. 

An acquisition function that can balance both exploration and exploitation is:
\begin{equation}
	\alpha(\mathbf{x^\star}) = -\frac{\left| y^\star \right|}{\sqrt{\mathrm{var}(y^\star) + \sigma^2}},
\end{equation}
where the negative sign ensures that a smaller absolute value of the mean leads to a larger acquisition function \cite{kapoor2010gaussian}. The rationale behind this choice is closely related to the t-test in hypothesis testing \cite{casella2002statistical}, which statisticians use to determine if numbers are sampled from a distribution with zero mean. Similarly, this acquisition function combines two criteria into one simple closed form. However, this choice has an obvious disadvantage. For any practical uses where a statepoint has dimensions $d \ge 2$, after sampling statepoints in both phases there will always be an estimated phase boundary in the GP interpolation where all statepoints will, by definition, have the same acquisition function value of $0$. This choice of acquisition function fails because of the degeneracy problem of treating all points on the estimated phase boundary as equally important, regardless of their different uncertainties. This can be seen visually in Fig. \ref{example}(d), where the acquisition function appears maximal along a single line. Therefore, to overcome this deficiency and choose the best trial statepoint, we remove this degeneracy with a small modification, and select as our acquisition function
\begin{equation}
	\alpha(\mathbf{x^\star}) = \frac{\sqrt{\mathrm{var}(y^\star) + \sigma^2}}{\left| y^\star \right| + \epsilon},
	\label{t_stat}
\end{equation}
where we compute the inverse of the original acquisition function, and add a small positive constant $\epsilon$ to the denominator to enforce numerical stability. The extra free parameter $\epsilon$ also has a physical meaning. Using the word ``margin" to refer to the distance between a trial point and the estimated phase boundary, $\epsilon$ essentially sets the cutoff of the margin's effect on the acquisition function as the margin approaches zero. This effectively sets an upper bound to the effect of the exploitation criterion. Empirically we find that a moderately small value between $0.05$ or $0.005$ suffices, depending on the desired exploration-exploitation trade-off. Thanks to the GP regression's analytical solution, we can efficiently find the trial statepoint with the largest acquisition function value by coding standard gradient-ascent optimization methods, or simply use packaged L-BFGS solvers \cite{liu1989limited}.

As a concrete example, we study a synthetic phase diagram with a ground truth phase boundary shown in Fig. \ref{example}(a). We warm up the algorithm with five randomly sampled statepoints and visualize all the intermediate values from the snapshot at iteration 20. Our algorithm uses GP interpolation to model the phase function, with the expected phase function shown in Fig. \ref{example}(b) and uncertainty shown in Fig. \ref{example}(c). The interpolation results are used to calculate the acquisition function value (Fig. \ref{example}(d)) according to Eq. \ref{t_stat}. Roughly, the algorithm can be understood as first taking the expected phase boundary and then selecting the next point to be the one on the expected boundary with maximum uncertainty.

\section{Batch Sampling}

\bigskip
\begin{figure*}
\begin{minipage}{\linewidth}
\begin{algorithm}[H]
  \caption{Active Learning Based Phase Diagram Sampling - Batch Sampling Algorithm}
  \label{algobatch}
  
\begin{flushleft}
   \textbf{Input:} {Randomly sampled initial dataset $\mathcal{D}_1 \defeq \left\{ \mathbf{x}_i, y_i \right\} ^n_{i=1}$ that contains both phases, maximum iteration $m$, batch size $n_b$.} 
\end{flushleft}

  \begin{algorithmic}
	\FOR {$t=1$ to $m$}
      \STATE {Fit a Gaussian Process to $\mathcal{D}_t$ and compute the acquisition function $\alpha(\mathbf{x})$ \;}
      \STATE {Initialize penalized acquisition function $\tilde{\alpha}_{t,0}(\mathbf{x}) = \alpha(\mathbf{x})$ \;}
      
      \FOR {$j=1$ to $n_b$}
          \STATE {Compute next trial point in the batch $\mathbf{x}^\star_{t,j} \leftarrow \argmax \tilde{\alpha}_{t,j-1}(\mathbf{x}))$ by gradient ascent method \;}
          \COMMENT {Select trial point by maximizing acquisition function}
          \IF {$\mu(\mathbf{x}_{t,j})=0$}
              \STATE {$r_j \leftarrow 1/L_\mu$ \;}
          \ELSE 
              \STATE {$r_j \leftarrow (\sigma-\sigma_0)/L_\sigma$ \;}
              \COMMENT {Compute penalizing radius $r_j$ for the last selected trial point}
          \ENDIF
          \STATE {Compute penalized acquisition function $\tilde{\alpha}_{t,j}(\mathbf{x}) \leftarrow \alpha_{t,0}(\mathbf{x}) \prod^j_{k=1} \varphi_{r_j}(\mathbf{x}, \mathbf{x}_k)$ \;}
      \ENDFOR
      
      \STATE {$\left\{ y^\star_{t,j}\right\} \leftarrow$ Parallel phase evaluations by experiment or simulation at $\left\{ \mathbf{x}^\star_{t,j} \right\}$ for $j=1$ to $n_b$ \;}
      \STATE {$\mathcal{D}_{t+1} \leftarrow \mathcal{D}_{t} \cup \left\{ (\mathbf{x}^\star_{t,1},y^\star_{t,1} ) \ldots (\mathbf{x}^\star_{t,n_b},y^\star_{t,n_b} ) \right\}$ \;}
      \COMMENT {Add the batch of observed statepoints in this round into the observed set}
    \ENDFOR
  \end{algorithmic}
\end{algorithm}
\end{minipage}
\end{figure*}

The previous sections describe the essential components for a fully functioning active learning framework proposing one maximally informative trial statepoint at each round, which we simply refer to as the ``sequential algorithm.'' In this section, we improve upon this method using batch sampling to exploit modern parallel architectures for both simulations and experiments. 

\begin{figure*}
\centering
\includegraphics[width=0.8\textwidth]{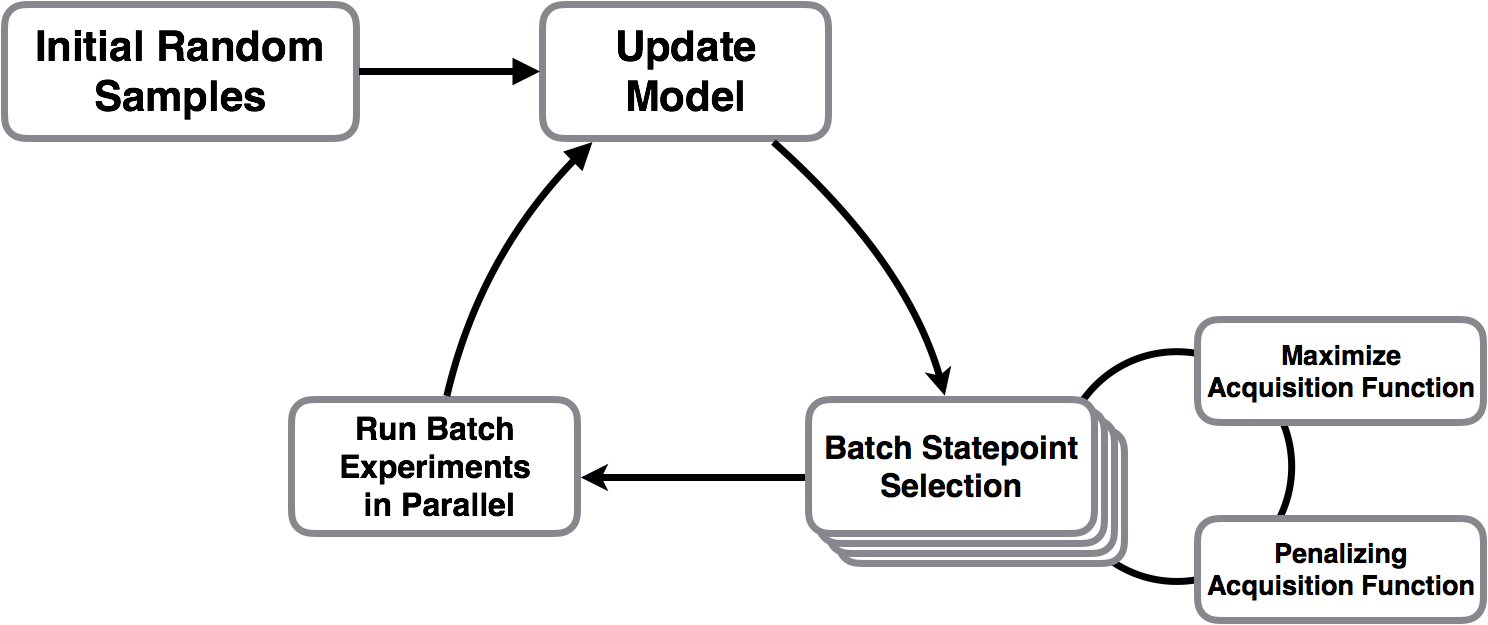}
\caption{A schematic that summarizes the batch active learning framework. Note that the problem of batch sampling is solved here by reducing it into a series of sequential statepoint selection problems in the penalization-maximization procedure.}
\label{schematics_batch}
\end{figure*}

While it is easy to achieve greedy Bayes optimality for our original sequential algorithm, it is highly nontrivial to achieve Bayes optimality for a batch sampling algorithm. Canonically, we would write down a multi-point acquisition function as a function of all the trial statepoints in the same batch, and solve the Markov decision process. Unfortunately, such a procedure is often impractical because of being analytically intractable or poor scaling with the dimension of the problem and the size of the batches \cite{gonzalez2016batch}. Various techniques and approximations have been developed in the Bayesian optimization community for optimizing the multi-point acquisition function \cite{kathuria2016batched,wang2016parallel,wu2016parallel,azimi2010batch}. However, many of the techniques are highly dependent on the specific functional form of the chosen acquisition function. Rather than directly optimizing the multi-point acquisition function, we adopt the recent idea of local penalization \cite{gonzalez2016batch} to develop a heuristic for batch sampling. The key insight is that the effect of evaluating a trial point in the optimal batch is just a local exclusion effect, preventing the algorithm from choosing other trial points in its proximity. By using good heuristics to penalize each trial point in the batch, we effectively reduce the problem of finding the optimal batch to a series of related sequential optimization problems. This framework of batch active learning is summarized in the schematic shown in Fig. \ref{schematics_batch}.

\begin{figure*}
\centering
\includegraphics[width=0.9\textwidth]{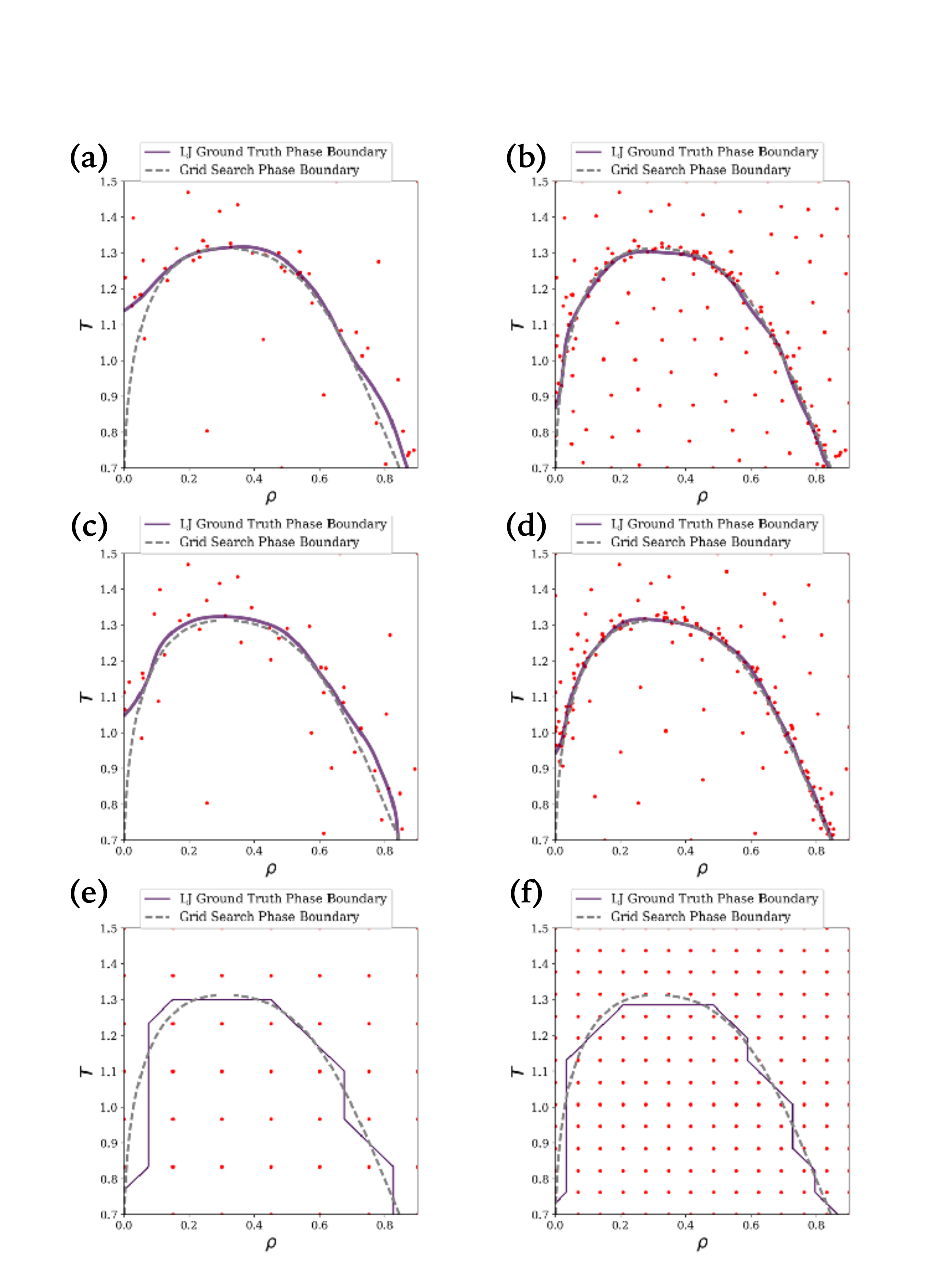}
\caption{Example application of active learning for a Lennard-Jones phase diagram as ground truth. The ground truth phase boundary is reconstructed with the liquid-vapor coexistence properties open data from NIST. Each column shows a different sampling algorithm, (a, b): grid search; (c, d): active learning sequential; (e, f): active learning batch with a batch size of 5. Each row shows snapshots with different numbers of sampled statepoints. Upper row: (a) 49 (7 by 7) samples; (c, e) 50 samples. Lower row: (b) 196 (14 by 14) samples; (d, f) 195 samples. In all subplots, the gray dashed curve shows the ground truth phase boundary as reference.}
\label{lj}
\end{figure*}

The aforementioned idea of local penalization is used to formally derive the ``maximization-penalization'' \cite{gonzalez2016batch} sampling procedure. In each sampling step, we approximate the optimal batch in a sequential fashion: first we choose the best trial point based on the current acquisition function and then penalize the acquisition function value around it. We iterate this process until the batch size is reached. Now we need only a proper formula that determines how large the exclusion regions should be.

Here we assume that $\mu$ and $\sigma$, our interpolation estimate's mean and standard deviation as functions of statepoint $\mathbf{x}$, are both following the Lipschitz assumption \cite{strongin2013global,gonzalez2016batch,sohrab2003basic}:
\begin{eqnarray}
	\| \mu(\mathbf{x_1}) - \mu(\mathbf{x_2}) \| \le L_\mu \| \mathbf{x_1} - \mathbf{x_2}\|, \\
    \| \sigma(\mathbf{x_1}) - \sigma(\mathbf{x_2}) \| \le L_\sigma \| \mathbf{x_1} - \mathbf{x_2}\|.
\end{eqnarray}
$L_\mu,L_\sigma$ are known as Lipschitz constants. They quantify the smoothness of our estimates, and can be easily estimated by sampling the largest global gradient magnitude for these functions as shown in \cite{gonzalez2016batch}.

To find the proper penalizing radius, first we notice that our original sequential algorithm tends to sample from two types of trial points -- those with high uncertainty and those lying on the estimated phase boundary. In both cases, the acquisition function of the evaluated trial point and its neighborhood decreases after evaluation. For the former type, the effect of sampling that particular trial point is to reduce the uncertainty of that region, making its neighborhood less likely to yield a high acquisition function later. Thus, the evaluation penalizes all points within an area around this trial statepoint at $\mathbf{x}_j$. We define $B_j = \left\{ \mathbf{x} : \| \mathbf{x}-\mathbf{x}_j\| \le r_j \right\}$ as the $d$-dimensional area around a selected trial point $\mathbf{x}_j$ with a radius of $r_j$, where the penalizing radius is given by $r_j = (\sigma - \sigma_0)/{L_\sigma}$. The rationale behind this formula is based on the Lipschitz assumption that the interpolation uncertainty $\sigma(\mathbf{x})$ cannot change faster than the rate of $L_\sigma$. Similarly, we can argue that for the latter type, the phase evaluation of a trial point would change its expected phase label from $0$ to $\pm 1$, shifting the estimated phase boundary accordingly and reducing the likelihood that its neighborhood will still contain a new phase boundary. Thus, for any point to be a potential candidate again (i.e. to be in the new estimated phase boundary where $\mu=0$), it must be as far from the last trial point $\mathbf{x}_j$ as $r_j=(1- \left| \mu(\mathbf{x}_j)\right|) / {L_\mu} = 1/L_\mu$.

Next, for every already-selected batch element, the penalizing multiplier to the acquisition function is given by
\begin{eqnarray}
    \varphi_{r_j}(\mathbf{x}, \mathbf{x}_j) &= 1 - \mathrm{Prob}(\mathbf{x}\in B_{r_j}(\mathbf{x}_j)) \\
    &= \mathrm{Prob}(r_j \le \|\mathbf{x}-\mathbf{x_j}\|),
\end{eqnarray}
which can be expressed in a standard Gauss error function following the treatment in \cite{gonzalez2016batch}. The above analysis leads to our batch sampling algorithm, shown in Algorithm \ref{algobatch}.

Empirically, we also find that combining the two penalizing criteria into one by taking the geometric mean of the penalizing radius as the global penalizing radius yields a batch algorithm variant that is slightly more robust (because more weight is put on exploration than exploitation). We choose this variant of batch algorithm in example applications in the next section.

\section{Example Applications and Discussion}
We wrote and implemented custom modified software for our active learning algorithm forked from the open source package GPyOpt\cite{gpyopt2016} and its dependency GPy\cite{gpy2014}. The acquisition function and the corresponding batch sampling method were added to GPyOpt as plugin modules. We tested our proposed sequential and batch algorithm on a 2D Lennard-Jones (LJ) system phase diagram. The phase diagram, shown in Fig. \ref{lj} (gray dashed curve) as the ground truth phase boundary, was compiled from liquid-vapor coexistence equation of state data \cite{nist}. We considered the phase boundary between the pure liquid / gas phase and the coexistence phase, where the phase boundary is exactly the phase coexistence equation of state.

In this example, the parameter for the acquisition function $\epsilon$ is set to 0.01. For consistency, we started both of our sequential and batch active learning algorithms with the same five random samples and actively sampled 190 LJ system simulations for both. In the sequential setting, this corresponds to 190 iterations; while in batch setting we chose a batch size of five, corresponding to 38 iterations. We examined the behavior of our sequential algorithm by taking two snapshots, one early in the run with 50 samples, including five random warm-up samples, the other at the end with 195 samples. We compared 7-by-7 grid search results (Fig. \ref{lj}(a)) with our sequential algorithm (Fig. \ref{lj}(c)) and batch algorithm (Fig. \ref{lj}(e)). Similarly, we compared 14-by-14 grid search results (Fig. \ref{lj}(b)) with our sequential algorithm (Fig. \ref{lj}(d)) and batch algorithm (Fig. \ref{lj}(f)). We find that our algorithms sample mostly around the phase boundary and less densely in other areas. The interpolated phase boundary is very close to the ground truth for both our algorithms, in the sense that our GP interpolation model can reconstruct a smooth phase boundary.

\begin{figure}
\centering
\includegraphics[width=0.4\textwidth]{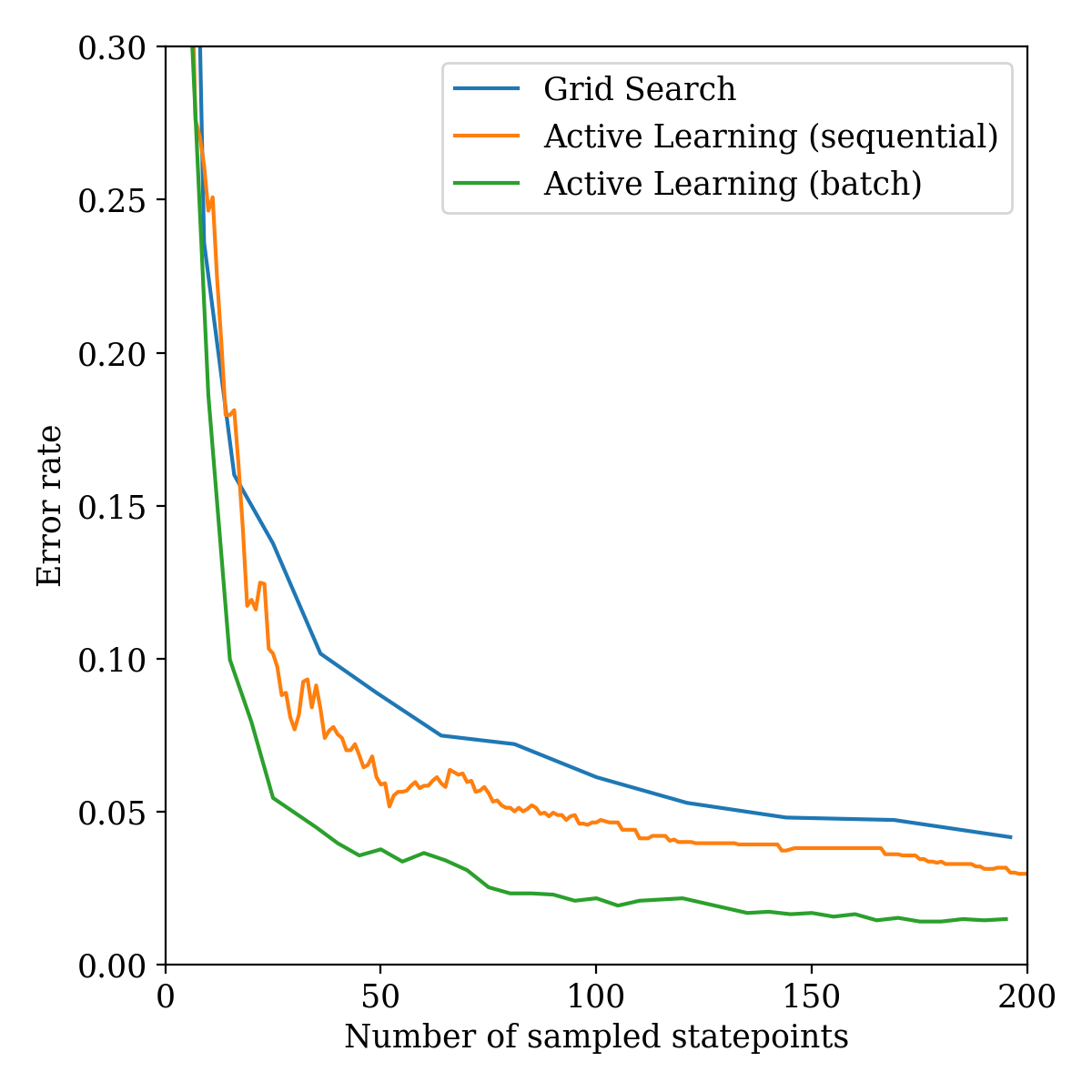}
\caption{Error rate vs. number of sampled statepoints. Both our batch and sequential algorithm outperform the grid search baseline by a large margin. The error rate of the reconstructed phase boundary is computed by constructing a fine grid of statepoints and calculating the percentage of statepoints incorrectly interpolated at every iteration of each algorithm. For grid search, the number of sampled statepoints are all perfect squares (e.g. 4-by-4, 5-by-5, etc). As anticipated, the error rate of active learning algorithms may occasionally increase and then correct itself very soon as more statepoints are sampled, because active learning algorithms always sample the statepoints that are most informative in the sense that they may incur large change in the estimated phase boundary.}
\label{rate}
\end{figure}

To quantitatively compare the performance of our active learning algorithms, we construct the error rate metric for phase diagram interpolation. We compute the error rate by first constructing a fine grid of statepoints and calculating the percentage of statepoints incorrectly interpolated. For the grid search method, we use the nearest neighbor interpolation model, i.e. an unobserved statepoint is interpolated with its nearest sampled statepoint's phase. In Fig. \ref{rate}, we show the error rates vs. number of sampled statepoints for our sequential and batch algorithms with the grid search method as a baseline. In both sequential and batch settings, our algorithms outperform grid search by a large margin, as shown in Fig. \ref{rate}. We note for that our algorithm, the error rate may rise occasionally as more experiments are sampled. This is because the algorithms always sample the next most informative statepoints, and the experiments' result might occasionally change the interpolation (i.e. being informative), causing large shifts in phase boundary interpolation. Our algorithms are robust enough to quickly correct such instabilities as a few more statepoints are sampled and continue improve the estimate of the phase boundary.

\section{Conclusion}
We proposed and demonstrated an active learning approach towards efficient phase diagram sampling. By proper choice of an acquisition function, our sequential sampling strategy outperforms conventional grid search sampling methods. Also, we generalized the local penalization techniques from Bayesian optimization research to our active learning framework, enabling the design of batch sampling to better take advantage of parallel testing available in most real world settings. We demonstrated our approach using a synthetic example and an example application in soft matter research. To extend our algorithms to the cases where more than two phases are present, we can reduce the problem to multiple one-vs-the-rest binary problems as discussed in \cite{kapoor2010gaussian,rasmussen2006gaussian}. In this reduction strategy, we can solve an $N$-phase active learning problem by solving $N$ binary problems in which we query each one of the possible phase's boundary with all other phases. Although our choice of example applications include only simulated data, our proposed active learning framework is equally compatible with experimental explorations of phase diagrams. We believe our method can greatly reduce the monetary cost and time for   material discovery and accelerate key applications such as complex material design, where conventional phase diagram sampling techniques impose tight bottlenecks.

We thank Jiachen (Cupjin) Huang for helpful discussions on the Lipschitz continuity. This work was supported as part of the Center for Bio-Inspired Energy Science (CBES), an Energy Frontier Research Center funded by the U.S. Department of Energy, Office of Science, Basic Energy Sciences under Award \# DE-SC0000989, and also by a Simons Investigator award from the Simons Foundation to S.C.G.  Computational resources and services were supported by Advanced Research Computing at the University of Michigan, Ann Arbor.

%--BIBLIOGRAPHY--%
\bibliographystyle{unsrt}
\bibliography{x}
\end{document}